\documentclass[a4paper]{jpconf}
\usepackage{graphicx}
\usepackage{amsmath}
\begin{document}
\title{Squeezed Thermal State Representation of the Inflaton and Particle Production in Bianchi type-I Universe}

\author{Karam Chand}

\address{Department of Physics, Malaviya National Institute of Technology Jaipur, 302017, India}

\ead{2015rpy9054@mnit.ac.in}

\begin{abstract}
In this study, we use  a single-mode squeezed thermal vacuum state formalism and examine the nature of a massive homogeneous scalar field, minimally coupled to the gravity in the framework of semiclassical gravity in the Bianchi type-I  universe. We have obtained an estimate leading solution to the semiclassical Einstein equation  for the Bianchi type-I universe shows, each scale factor in its respective direction obeys $ t^{\frac{2}{3}} $ power-law expansion. The mechanism of the nonclassical thermal cosmological particle production is also analyzed in the Bianchi type-I universe.\\
\\
\textbf{Keywords}: Squeezed Thermal state; Inflaton; Particle Production; Bianchi type-I Universe.
\end{abstract}

\section{Introduction}

The Standard Big Bang ($SBB$) model is an  achievement of the $ 20^{th} $ century that can confidently characterize the formation of the universe at a large scale after the primordial
nucleosynthesis. Even though it's a great success, the $SBB$ model had several longstanding unresolved aspects, some of them are the Horizon problem, the Monopole problem and the Flatness problem.    Cosmological inflationary model [1],  provides  satisfactory solutions to all these problems. At the present time, the inflationary epoch has many versions [1-3] of the cosmological inflationary model, called the inflationary-universe scenario.  The concept behind the cosmological inflationary model  is that  a period of accelerated expansion of space lasted from $10^{-36}$ seconds after the Big Bang in the early stage of the universe. The inflationary era can be well understood by consideration of a slow-roll massive inflaton [1, 3, 4], where the energy density of massive inflaton is entirely stored in the form of its potential energy density because  the contribution of the kinetic energy is very small  in the entire inflation period. The universe becomes cold after the inflationary era. Therefore, the temperature of the universe was not enough to start the primordial nucleosynthesis so,   the universe was empty in terms of various kinds of particles.  A particle production mechanism is needed to explain the decay of a massive inflaton into different types  of particles which is responsible for the reheating of the universe through thermalization. Near the potential minima a quasi-periodic motion of the massive inflaton with slowly decreasing amplitude can reveal the decay of massive inflaton into a particle shower [5-7]. Hence, at the end of the inflationary era, a significant amount of particles produced which helped to repopulate the universe with radiation and matter. The created particles moved freely and start  collisions among themselves. The temperature of the universe is enhanced due to the collisions of the created  particles and the decay of massive inflaton, therefore, the universe becomes hot again which was later known as the hot Big Bang model and the primordial nucleosynthesis mechanism triggered.  Therefore,  the oscillating phase of a massive  inflaton and its related aspects plays a key role to describe particle production and further formation  of the early universe.

Many of the cosmological inflation paradigms is based on the Friedmann-Robertson-Walker (FRW) universe model [8-11] in  which model is isotropic and homogeneous. The FRW cosmological models provide a satisfactory solution to the flatness problem while the solution is not much clear about homogeneity and isotropy. Absolutely the FRW cosmological models provide the necessary condition to achieve the solution of the horizon problem but do not provide the sufficient condition to solve the homogeneity and isotropy problems. According to Rothman and Ellis [12], an anisotropic metric consideration can provide solutions to isotropic problems they also demonstrated that such problems can be isotropized and inflated in many ways using some general conditions.  Therefore, it is reported [13-14] that anisotropies and inhomogeneities  might have played a key role in the evolutionary history of the early universe. A satisfactory solution is provided by the isotropic model for the explanation of the universe evolution at later stages but near the singularity or near the plank scale this model does not provide a suitable description of the universe at a very early stage [15]. Einstein’s theory of relativity provides cosmological solutions near the plank scale (singularity) in which expansion is an anisotropic at initial stages but at later stage evolution the solution becomes isotropic and near the singularity the gravitational collapse solution becomes anisotropic (locally) [16-18]. To remove the existence of particle horizon and postulating specific initial conditions to investigate isotropic models, a variety of corrections and initial conditions have been provided by rigorous study of an anisotropic and inhomogeneous models of the universe. 

The Bianchi type-I cosmological model is the simplest one model among the anisotropic cosmological models. In this model
the metric is assumed as spatially homogeneous and  anisotropic  in which a mechanism of isotropization of the universe is examined through the path of time. In
difference to the $ FRW $ metric,  the Bianchi type-I metric
has three cosmic scale factors that expand individually in their corresponding direction. Hence, the expansion of the universe in the Bianchi type-I  model can be assumed as anisotropic expansion. Recently such models received much attention [13, 14, 17-21]. In the Bianchi's type-I cosmological model Futamase has examined the
effective potential [21].  Huang using an adiabatic approximation for massless field with arbitrary coupling to gravity  has assumed the fate of symmetry
in the Bianchi type-I cosmological model [22]. Berkin has studied
the effective potential  for scalar field having
arbitrary mass and coupling to gravity in tha Bianchi type-I universe [23].  Recently,  the Bianchi type-I cosmological model also examined the first order phase transitions in the early universe [24]. These results reveal that
Bianchi type-I  model may be helpful to examine the very early stages of the universe. Anisotropic models of the universe which become isotropic have been
considered several times [16]. These motivate the
study of anisotropic background cosmological models with scalar field possessing
the advantage of the FRW model [25]. The possibility of
the Bianchi type-I universe approaches to the isotropic model can be examined.
From anisotropic to isotropic transition, a damping mechanism is required. One of
the efficient damping mechanisms could be due to particle creation in an anisotropic
models. Therefore it would be useful to examine particle production in an anisotropic
cosmological model with nonclassical thermal inflaton, which could expect to generate
sufficient amount of particles to bring isotropy during the evolution procedure  of the universe.

Many works are available in literature  based on the $ FRW $ model in which quantum effects and quantum properties of the massive inflaton were studied [26-28] in the inflationary phase. These research works reveal that the results obtained are
quite different from the classical results. Such kinds of studies show that quantum
effects of the matter field play a key part to understand the inflationary phase and
its associated phenomenon.
Some of the recent works got enough attention among the cosmologists for example works relating quantum optics with cosmology where nonclassical state representation such as coherent [29] and squeezed states [30, 31] for a massive inflaton are quite a beneficial to describe  quantum effects in the  cosmological model. The first plan  of adoption of the squeezed and coherent states for studying quantum effects and quantum properties in cosmology like cosmological particle production introduced by Grishchuk  and Sidorov [32, 33]. According to Gasperini and Giovannini [34] in the cosmological models of pair production the entropy generation can be related with the squeezing parameter. The inflationary cosmology studied by Albrecht et.al. [35]  in  light of squeezed states and so on [36-38]. A variety of cosmological problems can be solved by applying the existing physical as well as mathematical information from these states [39-47]. In such type of studies both squeezed and coherent states are assumed at absolute zero temperature. However, in the modern cosmology thermal properties of these states are an interesting research topic. Thermal counterparts of coherent and squeezed states are recognized as a coherent  thermal state [49-54] and squeezed thermal  state [50, 51, 55-58]. Thermal properties of the nonclassical states play an important role  to understand the thermal history and evolution of the early-stage universe [59-65]. According to cosmological perspective, consideration of thermal effects in flat FRW universe has many significance. Consequently, it motivates to study the thermal and quantum mechanical effects of a massive inflaton in  coherent and squeezed thermal states formalism in cosmology.

 In  this research article, we use single-mode squeezed thermal vacuum state ($STVS$) formalism and examine the nature of inflaton field, minimally coupled to the gravity  in the anisotropically expending universe which is determined by the  Bianchi type-I  universe in the framework of the semiclassical theory of gravity ($ SCTG $).  We  obtain an approximate leading solution to the $ SCEE $ for the Bianchi type-I universe. The mechanism of nonclassical thermal cosmological particle production is also analyze in the Bianchi type-I universe. We have used natural unit system $ G=c=\hbar=1 $  in the present study.
\section{Representations of  Single-Mode Squeezed Thermal Vacuum State}

In 1975 thermofield dynamics were developed by Takahashi and Umezawa in which the thermofield dynamics based on temperature dependent vacuum states [66-71]. The thermofield dynamics introduced that the expectation value of a mixed state at nonzero temperature is obtained by an equivalent calculation with a pure state. This is found by introducing an imaginary field which is a similar appearance to the natural real field. Thus the most significant aspect of thermofield dynamics is Bogoliubov canonical transformations that generate a theory from zero to finite temperature.  Hence, in this approach, a tilde space is needed in addition to the usual Hilbert space. By use of the Hilbert and the tilde space, we determine a
direct product space. In the Hilbert space, every operator and state  has the
corresponding equivalent part in the tilde space. Thermofield dynamics generate a
thermal operator $ \Gamma(\rho) $ based on  the Hilbert and tilde system,  which is invariant under the
tilde conjugation i.e., $ \tilde{\hat\Gamma}(\rho) = \hat{\Gamma}(\rho) $ [66-70]. \\

Hence, in thermofield dynamics, a temperature dependent  vacuum state   is described as [57, 72]
\begin{equation}\label{eq2.1a}
\vert \rho\rangle = \hat{\Gamma}(\rho)\vert0,\tilde{0}\rangle,
\end{equation}
where $ \rho $ is the thermal parameter, $ \vert 0,\tilde{0}\rangle $ represents temperature-independent  two-mode vacuum state and $ \hat{\Gamma}(\rho) \equiv \exp[\rho(\hat{{\cal A}}^{\dagger} \tilde{\hat{{\cal A}}}^{\dagger}-\hat{{\cal A}}\tilde{\hat{{\cal A}}})] $  stands for the  thermal operator. Here, Hilbert space creation and annihilation introduced by  $ \hat{{\cal A}}^{\dagger}$ and   $ \hat{{\cal A}} $  and  tilde space  creation and annihilation    operators described by $\tilde{\hat{{\cal A}}}^{\dagger}$  and $ \tilde{\hat{{\cal A}}}$. These  operators  are  obeying the  commutation relations $[\hat{{\cal A}}, \hat{{\cal A}}^{\dagger}]$ = $ [\tilde{\hat{{\cal A}}}, \tilde{\hat{{\cal A}}}^{\dagger}]= 1  $ in the Hilbert and tilde space respectively. By the appropriate action of thermal operator  on $ \hat{{\cal A}} $ and $ \hat{{\cal A}}^\dagger$, we obtain [57, 72]

\begin{equation}\label{eq2.3}
\hat{\Gamma}^{\dagger}(\rho)\ \hat{{\cal  A}}\ \hat{\Gamma}(\rho) =  \hat{{\cal A}}\cosh\rho+\tilde{\hat{{\cal A}}}^{\dagger}\sinh\rho ,
\end{equation}
and
\begin{equation}\label{eq2.4}
\hat{\Gamma}^{\dagger}(\rho)\ \hat{{\cal A}}^{\dagger}\ \hat{\Gamma}(\rho) =  \hat{{\cal A}}^{\dagger}\cosh\rho+\tilde{\hat{{\cal A}}}\sinh\rho.   
\end{equation}  
 Now the mean number of  produced particles in thermal vacuum state can be examined as: 
\begin{eqnarray}\label{eq2.5}
{\cal N} = \langle\rho\vert\hat{{\cal A}}^{\dagger}\hat{{\cal A}}\vert\rho\rangle = \sinh^2\rho.
\end{eqnarray} 
The thermofield dynamics is an effective approach to governing thermal properties of nonclassical  states. The main 
significant aspect of thermofield dynamics is Bogoliubov canonical transformations that generate a theory from zero to finite temperature. Therefore, 
 squeezed vacuum state with thermal effect is generated by operating
the thermal operator to the vacuum first and then followed by squeeze operator $ \hat{S}(r_{s},\theta) $
[73-75] it is described by

\begin{equation}\label{eq2.1}
\vert\varsigma, \rho\rangle_{\textit{STVS}} = \hat{S}(r_{s},\theta)\vert\rho\rangle,
\end{equation}
from (1)
\begin{equation}\label{eq2.7}
\vert\varsigma, \rho\rangle_{\textit{STVS}}=\hat{S}(r_{s},\theta)\hat{\Gamma}(\rho)\vert0,\tilde{0}\rangle,
\end{equation}

where squeezing  operator $\hat{S}(r_{s},\theta)  $ [75] takes the following form:

\begin{equation}\label{eq2.8}
\hat{S}(\vartheta)=\exp{1\over2}\left(\vartheta^{*}\hat{{\cal A}}^2-\vartheta\hat{{\cal A}}^{{\dagger}2}\right),
\end{equation}
 where $ \vartheta = r_{s}\ e^{i\theta}$. Its action on the creation  and annihilation  operators describes the following
  Bogoliubov canonical transformation 
\begin{align}\label{eq2.9}
\begin{aligned}
\hat{S}^{\dagger} \ \hat{{\cal A}}^{\dagger}\ \hat{S}= \hat{{\cal A}}^{\dagger}\ \cosh r_{s} -\hat{{\cal A}} \ e^{-i\theta}\sinh r_{s},
\\
\hat{S}^ {\dagger}\ \hat{{\cal A}}\ \hat{S} = \hat{{\cal A}}\ \cosh r_{s} -\hat{{\cal A}}^{\dagger} \ e^{i\theta}\sinh r_{s}.
\end{aligned} 
\end{align} 
In the present article,  the Caves's method [39] is used   to generate $ STVS $.
\section{Dynamics of Inflaton and SCEE in the Bianchi type-I  Metric}

Let us consider the Bianchi type-I universe which is spatially homogeneous and anisotropic. Therefore, the line element for such type of universe is given as:

   \begin{equation}\label{eq3.1}
   ds^2=-dt^2+{\cal Z}_{1}^2(t)dx^2+{\cal Z}_{2}^2(t)dy^2+{\cal Z}_{3}^2(t)dz^2,
   \end{equation}
where $ {\cal Z}_{1}(t) $, $ {\cal Z}_{2}(t) $ and $ {\cal Z}_{3}(t)$ are the cosmic scale factors describing the expansion of the universe in three spatial directions respectively and $t$ is the  physical time. 
The Bianchi type-I metric is a generalization of the FRW metric which is spatially anisotropic. All the three cosmic scale
factors i.e., $ {\cal Z}_{1}(t) $, $ {\cal Z}_{2}(t) $ and $ {\cal Z}_{3}(t) $  are derived via Einstein’s field equations. \vspace{1mm}

We   consider minimally coupling of  massive inflton to gravity therefore the background metric (\ref{eq3.1}) obeying the following relation

\begin{equation}\label{eq3.2}
(g^{\alpha\beta}\nabla_{\alpha}
 \nabla_{\beta}-m^{2})\Phi(x, t)= 0,
\end{equation}
where $  g^{\alpha\beta}$, $ \nabla_{\alpha} $, $ \Phi(x,t) $ and 'm'  represents the metric tensor, covariant derivative, scalar field and mass of the scalar field respectively. Now, the Lagrangian density of a massive inflton can be expressed as: 
\begin{equation}\label{eq3.3}
  {\cal L} =
-{1\over2}\sqrt{-g}(g^{\alpha\beta}\partial_\alpha\Phi\partial_\beta\Phi+m^2\Phi^2),
\end{equation}
where  $ g=\vert g_{\alpha\beta}\vert $  represents the metric tensor determinant.
In the present study we consider the scalar field  spatially homogeneous, i.e., $\Phi(x, t)=\Phi(t)$, therefore, classically a massive inflaton follows the Klein-Gordon equation given by Eq. (\ref{eq3.1}) and (\ref{eq3.3}) as:
 \begin{equation}\label{eq3.4}
\ddot{\Phi}(t)+\sum_{i=1}^{3} \left(  {\dot{{\cal Z}_{i}}(t)\over {\cal Z}_{i}(t)}\right) \dot{\Phi}(t)+m^2\Phi(t)=0,
\end{equation}
where dots stand for cosmic time $ t $ derivatives.
Hubble parameter $ {\cal H} =\frac{1}{3}\biggr(\frac{\dot{{\cal Z}_{1}}(t)}{{\cal Z}_{1}(t)}+\frac{\dot{{\cal Z}_{2}}(t)}{{\cal Z}_{2}(t)}+\frac{\dot{{\cal Z}_{3}}(t)}{{\cal Z}_{3}(t)}\biggr) $  plays the role of friction term in field dynamics in the spatial directions $ x $, $ y $, and $ z $,
respectively. Now  massive inflaton can be quantized, by canonical quantization rules, in a way consistent with the equation of motion, the conjugate momenta corresponding to  $\Phi$, is given as - $\Pi_{\Phi}={\partial {\cal L}}/{\partial{\dot{\Phi}}}$.\\

A time-dependent harmonic oscillator model can be used to describe the evolution of the minimally coupled massive inflaton to gravity in the Bianchi type-I universe, with a generalized Hamiltonian obtained by using
Eq.(\ref{eq3.3})   in the following relation
 \begin{equation}\label{eq3.6}
 H=\Pi_{\Phi}\dot{\Phi}-{\cal L}, 
 \end{equation}
 hence the Hamiltonian for the massive inflaton is determined as
 \begin{equation}\label{eq3.7}
 H_{m}=\frac{\Pi^2_{\Phi}}{2{\cal Z}_{1}(t){\cal Z}_{2}(t){\cal Z}_{3}(t)}+\frac{1}{2}{\cal Z}_{1}(t){\cal Z}_{2}(t){\cal Z}_{3}(t)m^2\Phi^2(t).
 \end{equation} 
Here, we assumed the minimal coupling between massive inflaton and the gravity Therefore, the classical Friedmann equation  can be defined as:
\begin{equation}\label{eq3.8}
{\dot{{\cal Z}}_{1}(t)\over {\cal Z}_{1}(t)}{\dot{{\cal Z}}_{2}(t)\over {\cal Z}_{2}(t)}+{\dot{{\cal Z}}_{2}(t)\over {\cal Z}_{2}(t)}{\dot{{\cal Z}}_{3}(t)\over {\cal Z}_{3}(t)}+{\dot{{\cal Z}}_{1}(t)\over {\cal Z}_{1}(t)}{\dot{{\cal Z}}_{3}(t)\over {\cal Z}_{3}(t)}={8\pi\over3}{{\cal T}_{00}\over {\cal Z}_{1}(t){\cal Z}_{2}(t){\cal Z}_{3}(t)},
\end{equation}
where ${\cal{T}}_{00} = {\cal Z}_{1}(t){\cal Z}_{2}(t){\cal Z}_{3}(t)
\left({1\over2}\dot{\Phi}^2+{1\over2} m^2\Phi^2\right)$ represents the time-time (00) component of the matter stress  tensor for the massive inflaton. In the cosmological context, the classical description of the Friedmann 
equation (\ref{eq3.8}) means that the Hubble parameter, ${ \cal H}_{i} =\frac{\dot{{\cal Z}}_{i}(t)}{{\cal Z}_{i}(t)} $, is calculated by
the energy density of the dynamically evolving massive inflaton as mentioned by  Eq.(\ref{eq3.4}).

\section{SCEE and Massive Inflation in Nonclassical Thermal  State}

Most inflationary universe models are established on the classical gravity with
Einstein equations and the unquantized (classical) inflaton field on the $ FRW $
metric.
To understand the early universe at a deeper level it is needed both classical
 Einstein equations and inflaton field are to be quantized. Despite
many efforts in the last few decades, there does not yet exist a consistent theory
of quantum gravity. However quantum effect and quantum
properties of the inflaton plays a very significant role to solve many cosmological
problems [39-47]. Therefore, for the proper explanation of a cosmological model,
we apply an approximation on Einstein field equations are known as a semiclassical approximation or $ SCEE $ [76].
Such an approximation of the Einstein field equations are also expected to be valid
where the quantum gravity effects are negligible [77, 78].   \\
In semiclassical approximation,  the Einstein field equation takes the following form 

\begin{equation}\label{eq4.1}
G_{\alpha\beta}= {8\pi} \langle \Psi \vert\hat{{\cal T}}_{\alpha\beta}\vert\Psi\rangle.
\end{equation}
Eq. (16) is known as $ SCEE $ and where the left side of the above  equation is a function of the metric: $ G^{\alpha\beta} $ is a four-dimensional Einstein curvature tensor representing the curvature of space-time, computed from the metric defined as: $ \Re_{\alpha\beta}-\frac{1}{2}\Re\ g_{\alpha\beta} $ and the right side angle brackets determine the expectation value of the matter stress tensor operator in the suitable quantum state $ \vert \Psi\rangle $ which is obeying the  Schr\"{o}dinger equation

\begin{equation}\label{eq4.2}
i\frac{\partial}{\partial t}\vert\Psi\rangle =\hat{ H}_{m}\vert\Psi\rangle,
\end{equation}
where $ \hat{{ H}}_{m} $ stands for Hamiltonian operator for the matter field.\\
 In the  $ SCTG $, the semiclassical Friedmann equation in the  context of the Bianchi type-I model  can be determined  as
\begin{equation}\label{eq4.3}
{\dot{{\cal Z}}_{1}(t)\over {\cal Z}_{1}(t)}{\dot{{\cal Z}}_{2}(t)\over {\cal Z}_{2}(t)}+{\dot{{\cal Z}}_{2}(t)\over {\cal Z}_{2}(t)}{\dot{{\cal Z}}_{3}(t)\over {\cal Z}_{3}(t)}+{\dot{{\cal Z}}_{1}(t)\over {\cal Z}_{1}(t)}{\dot{{\cal Z}}_{3}(t)\over {\cal Z}_{3}(t)}=\frac{8\pi}{ {\cal Z}_{1}(t){\cal Z}_{2}(t){\cal Z}_{3}(t)}\langle\hat{ H}_{m}\rangle_{\Psi},
\end{equation}
where $ H_{m} $ is given by Eq.(14)\\
 
The Fock space of  Hamiltonian (\ref{eq3.7}) was constructed in [79]
\begin{equation}\label{eq4.5}
\hat{{\cal A}}^{\dagger}(t)\hat{{\cal A}}(t)\vert n,\Phi,t\rangle = n\vert n,\Phi,t\rangle,
\end{equation}
so, in the Bianchi type-I model
the annihilation and creation operators can be written in the term of fields operators as
\begin{align}\label{eq4.6}
\hat{{\cal A}}(t)
&=\Phi^*(t)\hat{\Pi}-{\cal Z}_{1}(t){\cal Z}_{2}(t){\cal Z}_{3}(t)\dot{\Phi}^*(t)\hat{\Phi},\nonumber\\
\hat{{\cal A}}^{\dagger}(t) &=\Phi(t)\hat{\Pi}-{\cal Z}_{1}(t){\cal Z}_{2}(t){\cal Z}_{3}(t)\dot{\Phi}(t)\hat{\Phi}.
\end{align}
From Eq. (\ref{eq4.6})  we can be calculated the field operators in the following form
\begin{align}\label{eq4.7}
\hat{\Phi}= \frac{1}{i}(\Phi^*(t)\hat{{\cal A}}^{\dagger}(t)-\Phi(t) \hat{{\cal A}}(t))
\end{align}
\begin{align}\label{eq4.8}
\hat{\Pi}=i{\cal Z}_{1}(t){\cal Z}_{2}(t){\cal Z}_{3}(t)(t)(\dot{\Phi}(t)\hat{{\cal A}}(t)-\dot{\Phi^*}(t)\hat{{\cal A}}^{\dagger}(t)).
\end{align}
Next, we take the massive inflaton $ \Phi $ and its conjugate $ \Pi $  in single-mode $ STVS $  by substituting the
number state $ \vert n,\Phi,t\rangle = \vert \varsigma\rangle$ and calculated as

\begin{align}\label{eq4.9}
\langle\hat{\Phi}^2\rangle_{STVS}&=\biggr[(2(\sinh^2\rho\ \cosh^2r_{s}-2A\cos\theta+\cosh^2\rho\ \sinh^2r_{s})+1){\Phi}^* {\Phi}\nonumber\\
&-(K){\Phi}^{*2}
-(L){\Phi}^2\biggr],
\end{align}
and
\begin{align}\label{eq4.10}
\langle\hat{\Pi}^2\rangle_{STVS}&={\cal Z}_{1}^2(t){\cal Z}_{2}^2(t){\cal Z}_{3}^2(t)\biggr[(2(\sinh^2\rho\  \cosh^2r_{s}-2A\cos\theta+\cosh^2\rho\ \sinh^2r_{s})\nonumber\\
&+1)\dot{\Phi}^*\dot{\Phi}-(K)\dot{\Phi}^{*2}
-(L)\dot{\Phi}^2\biggr],
\end{align}
where 
\begin{align}\label{eq4.11}
K=B \cosh^2r_{s}-Ce^{-i\theta}\sinh^2\rho-C e^{-i\theta}\cosh^2\rho+B \sinh^2r_{s} \ e^{-2i\theta},
\end{align}
\begin{align}\label{eq4.12}
L=B \cosh^2r_{s}-Ce^{i\theta}\sinh^2\rho-Ce^{i\theta}\cosh^2\rho
+B\sinh^2r_{s}\ e^{2i\theta},
\end{align}
\begin{align}\label{eq4.13}
 A=\sinh r_{s} \  
\cosh r_{s}\ \sinh\rho\ \cosh\rho,
\end{align}
and
\begin{align}\label{eq4.14}
B=\sinh\rho\ \cosh\rho, C =\sinh r_{s} \  \cosh r_{s}.
\end{align}
Now,  we calculate the Hamiltonian  in single-mode $ STVS $ using Eqs. (\ref{eq4.9}) and (\ref{eq4.10} ) obtained as

\begin{align}\label{eq4.15}
\langle\hat{H}_m\rangle_{STVS}=&{\cal Z}_{1}(t){\cal Z}_{2}(t){\cal Z}_{3}(t)\biggr[\biggr(\sinh^2\rho\ \cosh^2r_{s}-2A\cos\theta+\cosh^2\rho\ \sinh^2r_{s}\nonumber\\
&+{1\over2}\biggr)(\dot{\Phi}^*\dot{\Phi}+m^2\Phi^*\Phi)\nonumber\\
&-{1\over2}(K)(\dot{\Phi}^{*2}+m^2\Phi^{*2})\nonumber\\
&-{1\over2}(L)(\dot{\Phi}^2+m^2\Phi^2)\biggr],
\end{align} 
where $ K $ and $ L $ are given in Eqs. (\ref{eq4.11}), (\ref{eq4.12})  respectively and  
the semiclassical Friedmann equation in single-mode $ STVS $ takes the following form

\begin{align}\label{eq4.16}
{\dot{{\cal Z}}_{1}(t)\over {\cal Z}_{1}(t)}{\dot{{\cal Z}}_{2}(t)\over {\cal Z}_{2}(t)}+{\dot{{\cal Z}}_{2}(t)\over {\cal Z}_{2}(t)}{\dot{{\cal Z}}_{3}(t)\over {\cal Z}_{3}(t)}+{\dot{{\cal Z}}_{1}(t)\over {\cal Z}_{1}(t)}{\dot{{\cal Z}}_{3}(t)\over {\cal Z}_{3}(t)}&= 8\pi \biggr[\biggr(\sinh^2\rho\ \cosh^2r_{s}-2A\cos\theta+\cosh^2\rho\ \nonumber\\&\times\sinh^2r_{s}+{1\over2}\biggr)(\dot{\Phi}^*\dot{\Phi}+m^2\Phi^*\Phi)\nonumber\\
&-{1\over2}(K)(\dot{\Phi}^{*2}+m^2\Phi^{*2})\nonumber\\
&-{1\over2}(L)(\dot{\Phi}^2+m^2\Phi^2)\biggr].
\end{align}
In  Eq.(\ref{eq4.16}) $ \Phi^* $ and  $ \Phi $ obey the Wronskian boundary condition which is described as
\begin{equation}\label{eq4.17}
\dot{\Phi}^{*}(t)\Phi(t)-\Phi^*(t)\dot{\Phi}(t)={i\over {\cal Z}_{1}(t){\cal Z}_{2}(t){\cal Z}_{3}(t)}.
\end{equation}
Next, we solve analytically the self-consistent $ SCEE $ Eq. (\ref{eq4.16}) corresponding to  single-mode  $ STVS $ for which, we can be  transformed the solution of a massive inflaton  in the following form
\begin{equation}\label{eq4.18}
\Phi(t)={1\over [{\cal Z}_{1}(t){\cal Z}_{2}(t){\cal Z}_{3}(t)]^{\frac{1}{2}}}\Omega(t),
\end{equation}
hence Eq.(\ref{eq3.4}) takes the following form 
\begin{align}\label{eq4.19}
\ddot{\Omega}(t)+\left(m^2 + \frac{1}{4}\sum_{i=1}^{3} \left(  {\dot{{\cal Z}_{i}}(t)\over {\cal Z}_{i}(t)}\right)^{2}-\frac{1}{2}\sum_{i\neq j=1}^{3} \left({\dot{{\cal Z}}_{i}(t)\over {\cal Z}_{i}(t)}{\dot{{\cal Z}}_{j}(t)\over {\cal Z}_{j}(t)}\right)  - \frac{1}{2}\sum_{i=1}^{3} {\ddot{{\cal Z}}_{i}(t)\over {\cal Z}_{i}(t)}\right)\Omega(t)=0.
\end{align}
 In the present study, our main focus is  the oscillatory phase of massive inflton in the matter-dominated era.  Therefore in the parameter regime massive inflaton obey the following inequality 
  
\begin{equation}\label{eq4.20}
m^2>\frac{1}{4}\sum_{i=1}^{3} \left(  {\dot{{\cal Z}_{i}}(t)\over {\cal Z}_{i}(t)}\right)^{2}-\frac{1}{2}\sum_{i\neq j=1}^{3} \left({\dot{{\cal Z}}_{i}(t)\over {\cal Z}_{i}(t)}{\dot{{\cal Z}}_{j}(t)\over {\cal Z}_{j}(t)}\right)  - \frac{1}{2}\sum_{i=1}^{3} {\ddot{{\cal Z}}_{i}(t)\over {\cal Z}_{i}(t)},
\end{equation}
therefore, an    oscillatory solution of the massive inflaton in the matter-dominated era can be  described as [80]
\begin{equation}\label{eq4.21}
\Omega(t)={1\over\sqrt{2\sigma(t)}}\exp(-i\int \sigma(t)dt),
\end{equation}
with
\begin{align}\label{eq4.22}
\sigma(t) = & \biggr( m^2-\frac{1}{4}\sum_{i=1}^{3} \left(  {\dot{{\cal Z}_{i}}(t)\over {\cal Z}_{i}(t)}\right)^{2}-\frac{1}{2}\sum_{i\neq j=1}^{3} \left({\dot{{\cal Z}}_{i}(t)\over {\cal Z}_{i}(t)}{\dot{{\cal Z}}_{j}(t)\over {\cal Z}_{j}(t)}\right)  - \frac{1}{2}\sum_{i=1}^{3} {\ddot{{\cal Z}}_{i}(t)\over {\cal Z}_{i}(t)}\nonumber\\&
+{3\over4}\left({\dot{\sigma}(t)\over \sigma(t)}\right)^2-{3\over2}
{\ddot{\sigma}(t)\over \sigma(t)}\biggr)^{\frac{1}{2}}.
\end{align}
Using Eqs.(\ref{eq4.18}), (\ref{eq4.20}) and  (\ref{eq4.21}) and    by calculating $\Phi,\Phi^*,\dot{\Phi}$ and $\dot{\Phi}^*$, the semiclassical Friedmann equation (\ref{eq4.16}) can be   obtained as

\begin{align}\label{eq4.23}
{\cal Z}_{1}(t){\cal Z}_{2}(t){\cal Z}_{3}(t)&= \frac{8\pi}{2\sigma\biggr({\dot{{\cal Z}_{1}(t)}\over {\cal Z}_{1}(t)}{\dot{{\cal Z}_{2}(t)}\over {\cal Z}_{2}(t)}+{\dot{{\cal Z}_{2}(t)}\over {\cal Z}_{2}(t)}{\dot{{\cal Z}_{3}(t)}\over {\cal Z}_{3}(t)}+{\dot{{\cal Z}_{1}(t)}\over {\cal Z}_{1}(t)}{\dot{{\cal Z}_{3}(t)}\over {\cal Z}_{3}(t)}\biggr)} \biggr[\biggr(\sinh^2\rho\ \cosh^2r_{s}\nonumber\\&-2A\cos\theta+\cosh^2\rho\ \sinh^2r_{s}+{1\over2}\biggr)\biggr(\frac{1}{4}\sum_{i, j=1}^{3} \left({\dot{{\cal Z}}_{i}(t)\over {\cal Z}_{i}(t)}{\dot{{\cal Z}}_{j}(t)\over {\cal Z}_{j}(t)}\right)\nonumber\\&+\frac{3}{4}\sum_{i=1}^{3} \left(  {\dot{{\cal Z}_{i}}(t)\over {\cal Z}_{i}(t)}\right)\frac{\dot{\sigma}}{\sigma}+\frac{1}{4}\biggr(\frac{\dot{\sigma}}{\sigma}\biggr)^2 +m^2+\sigma^2\biggr)\nonumber\\
&-{1\over2}K\ e^{2i\sigma t}\biggr(\frac{1}{4}\sum_{i, j=1}^{3} \left({\dot{{\cal Z}}_{i}(t)\over {\cal Z}_{i}(t)}{\dot{{\cal Z}}_{j}(t)\over {\cal Z}_{j}(t)}\right)+\frac{3}{4}\sum_{i=1}^{3} \left(  {\dot{{\cal Z}_{i}}(t)\over {\cal Z}_{i}(t)}\right)\frac{\dot{\sigma}}{\sigma}\nonumber\\&-i\sigma\sum_{i=1}^{3} \left(  {\dot{{\cal Z}_{i}}(t)\over {\cal Z}_{i}(t)}\right)-\sigma^2 +m^2\biggr)\nonumber\\
&-{1\over2}L \ e^{-2i\sigma t}\biggr(\frac{1}{4}\sum_{i, j=1}^{3} \left({\dot{{\cal Z}}_{i}(t)\over {\cal Z}_{i}(t)}{\dot{{\cal Z}}_{j}(t)\over {\cal Z}_{j}(t)}\right)+\frac{3}{4}\sum_{i=1}^{3} \left(  {\dot{{\cal Z}_{i}}(t)\over {\cal Z}_{i}(t)}\right)\frac{\dot{\sigma}}{\sigma}\nonumber\\&+i\sigma\sum_{i=1}^{3} \left(  {\dot{{\cal Z}_{i}}(t)\over {\cal Z}_{i}(t)}\right)-\sigma^2 +m^2\biggr)\biggr].
\end{align}
Next, we solved perturbatively the above Eq. (37) in single-mode $ STVS $, further on using the following approximations, which corresponds to the
parametric resonance conditions:

\begin{align}\label{eq4.24}
\sigma_0(t) &=m;~~~~~~{\cal Z}_{10}(t)={\cal Z}_{10} t^{2/3}, {\cal Z}_{20}(t)={\cal Z}_{20} t^{2/3}, {\cal Z}_{30}(t)={\cal Z}_{30} t^{2/3}\nonumber\\
\dot{\sigma}_0(t) &=0;~~~~~~  \frac{{\dot{\cal Z}}_{10}(t)}{{\cal Z}_{10}(t)}=\frac{{\dot{\cal Z}}_{20}(t)}{{\cal Z}_{20}(t)}=\frac{{\dot{\cal Z}}_{30}(t)}{{\cal Z}_{30}(t)}=\frac{2}{3t}.
\end{align}
Now using  the above mentioned parametric  condition Eq.
(\ref{eq4.24}), the next order perturbative solution for cosmic scale factor ${\cal Z}_{1}(t)  $ is obtained as

\begin{align}\label{eq4.25}
{\cal Z}_{11}(t)=&\frac{6\pi}{{\cal Z}_{20}{\cal Z}_{30}}\biggr[\biggr(\sinh^2\rho\ \cosh^2r_{s}-2A\cos\theta+\cosh^2\rho\ \sinh^2r_{s}+{1\over2}\biggr)\biggr(1+\frac{1}{2m^2t^2}\biggr)\nonumber\\&-\frac{1}{2m^2t^2}\biggr(B\cosh^2r_{s}\ \cos(2mt)-C(\sinh^2\rho+\cosh^2\rho)\cos(\theta-2mt)\nonumber\\&+B\sinh^2r_{s}\ \cos2(\theta-mt) \biggr)-\frac{1}{mt}\biggr(B\cosh^2r_{s}\ \sin(2mt)-C(\sinh^2\rho\nonumber\\&+\cosh^2\rho)\sin(\theta-2mt)+B\sinh^2r_{s}\ \sin2(\theta-mt) \biggr)\biggr]m \ t^{\frac{2}{3}}.
\end{align}
In the same way, we  have obtained the
 the next order perturbation solution for cosmic scale factors ${\cal Z}_{2}(t)  $ and ${\cal Z}_{3}(t)  $ in single-mode $ STVS $ as

\begin{align}\label{eq4.26}
{\cal Z}_{21}(t)=&\frac{6\pi}{{\cal Z}_{10}{\cal Z}_{20}}\biggr[\biggr(\sinh^2\rho\ \cosh^2r_{s}-2A\cos\theta+\cosh^2\rho\ \sinh^2r_{s}+{1\over2}\biggr)\biggr(1+\frac{1}{2m^2t^2}\biggr)\nonumber\\&-\frac{1}{2m^2t^2}\biggr(B\cosh^2r_{s}\ \cos(2mt)-C(\sinh^2\rho+\cosh^2\rho)\cos(\theta-2mt)\nonumber\\&+B\sinh^2r_{s}\ \cos2(\theta-mt) \biggr)-\frac{1}{mt}\biggr(B\cosh^2r_{s}\ \sin(2mt)-C(\sinh^2\rho\nonumber\\&+\cosh^2\rho)\sin(\theta-2mt)+B\sinh^2r_{s}\ \sin2(\theta-mt) \biggr)\biggr]m \ t^{\frac{2}{3}},
\end{align}
 and

\begin{align}\label{eq4.27}
{\cal Z}_{31}(t)=&\frac{6\pi}{{\cal Z}_{10}{\cal Z}_{20}}\biggr[\biggr(\sinh^2\rho\ \cosh^2r_{s}-2A\cos\theta+\cosh^2\rho\ \sinh^2r_{s}+{1\over2}\biggr)\biggr(1+\frac{1}{2m^2t^2}\biggr)\nonumber\\&-\frac{1}{2m^2t^2}\biggr(B\cosh^2r_{s}\ \cos(2mt)-C(\sinh^2\rho+\cosh^2\rho)\cos(\theta-2mt)\nonumber\\&+B\sinh^2r_{s}\ \cos2(\theta-mt) \biggr)-\frac{1}{mt}\biggr(B\cosh^2r_{s}\ \sin(2mt)-C(\sinh^2\rho\nonumber\\&+\cosh^2\rho)\sin(\theta-2mt)+B\sinh^2r_{s}\ \sin2(\theta-mt) \biggr)\biggr]m \ t^{\frac{2}{3}}.
\end{align}
Where $ {\cal Z}_{11}(t) $, $ {\cal Z}_{21}(t) $ and $ {\cal Z}_{31}(t) $ represent the next order perturbative solution for the cosmic scale factor $ {\cal Z}_{1}(t) $ in the
$ x $, $ y $ and $  z$ directions respectively. 
 From the above three Eqs. (\ref{eq4.25}), (\ref{eq4.26}) and (\ref{eq4.27})   it follows that

\begin{align*}
{\cal Z}_{11}(t)\sim t^{\frac{2}{3}},
\end{align*}

\begin{align}
{\cal Z}_{21}(t)\sim t^{\frac{2}{3}},
\end{align}
and 
\begin{align*}
{\cal Z}_{31}(t)\sim t^{\frac{2}{3}}.
\end{align*}
Here, we observed that all cosmic scale factors obey the same power-law of expansion i.e., $ t^{\frac{2}{3}} $.
\section{ Thermal and Quantum Particle Production in the Bianchi type-I universe}

In this section,  using single-mode $ STVS $ formalism we analyze the mechanism of particle production due to oscillations of massive inflaton in the anisotropically expanding universe determined through the Bianchi type-I  metric in the framework of $SCTG$.  For this, first, we examine the Fock space which has a parameter based on the cosmological time $t$. Then at a later time $t$, the number of produced particles from the vacuum states at an initial time $t_0$ is given by
\begin{equation}\label{eq5.1}
{\cal N}_0(t,t_0) = \langle 0,\Phi,t_0)\vert\hat{{\cal N}}(t)\vert0,\Phi,t_0\rangle,
\end{equation}
where $ \hat{{\cal N}}(t)=\hat{{\cal A}}^{\dagger}(t)\hat{{\cal A}}(t) $, is a number operator and  the vacuum expectation value of the right-hand side of Eq. (43) can be computed as 
\begin{align}\label{eq5.2}
\langle\hat{{\cal N}}(t)\rangle
&=\Phi(t)\Phi^*(t)\langle\hat{\Pi}^2\rangle-{\cal Z}_{1}(t){\cal Z}_{2}(t){\cal Z}_{3}(t)\Phi(t)\dot{\Phi}^*(t)\langle\hat{\Pi}\hat{\Phi}\rangle\nonumber\\
&-{\cal Z}_{1}(t){\cal Z}_{2}(t){\cal Z}_{3}(t)\dot{\Phi}(t)\Phi^*(t)\langle\hat{\Phi}\hat{\Pi}\rangle+({\cal Z}_{1}(t){\cal Z}_{2}(t){\cal Z}_{3}(t))^2\dot{\Phi}(t)\dot{\Phi}^*(t)\langle\hat{\Phi}^2\rangle,
\end{align}
here $\langle\hat{\Pi}^2\rangle, \langle\hat{\Pi}\hat{\Phi}\rangle$, $\langle\hat{\Phi}\hat{\Pi}\rangle$ and $\langle\hat{\Phi}^2\rangle$ are  respectively obtained as
\begin{align}\label{eq5.3}
\langle\hat{\Pi}^2\rangle
&=({\cal Z}_{1}(t){\cal Z}_{2}(t){\cal Z}_{3}(t))^2\dot{\Phi}^*\dot{\Phi}~~;~~\langle\hat{\Pi}\hat{\Phi}\rangle={\cal Z}_{1}(t){\cal Z}_{2}(t){\cal Z}_{3}(t)\dot{\Phi}\Phi^*;\nonumber\\
\langle\hat{\Phi}\hat{\Pi}\rangle
&={\cal Z}_{1}(t){\cal Z}_{2}(t){\cal Z}_{3}(t)\Phi\dot{\Phi}^*~~;~~\langle\hat{\Phi}^2\rangle=\Phi^*\Phi,
\end{align}
now putting Eq.(\ref{eq5.3}) in  Eq.(\ref{eq5.2}), then we obtain
\begin{equation}\label{eq5.4}
{\cal N}_0(t,t_0) = ({\cal Z}_{1}(t){\cal Z}_{2}(t){\cal Z}_{3}(t))^2\vert\Phi(t)\dot{\Phi}(t_0)-\dot{\Phi}(t)\Phi(t_0)\vert^2,
\end{equation}
using the perturbative solution in the limit  $mt_{0}$, $mt >1$  the number of created particles  in  the vacuum states 
  can be calculated as [81] 
\begin{align}\label{eq5.5}
{\cal N}_0(t,t_0) & = {1\over4\sigma(t)\sigma(t_0)} {{\cal Z}_{1}(t){\cal Z}_{2}(t){\cal Z}_{3}(t)\over {\cal Z}_{1}(t_0){\cal Z}_{2}(t_0){\cal Z}_{3}(t_0)}
\biggr[{1\over4}\sum_{i= j=1}^{3} \left({\dot{{\cal Z}}_{i}(t)\over {\cal Z}_{i}(t)}{\dot{{\cal Z}}_{j}(t)\over {\cal Z}_{j}(t)}+{\dot{{\cal Z}}_{i}(t_{0})\over {\cal Z}_{i}(t_{0})}{\dot{{\cal Z}}_{j}(t_{0})\over {\cal Z}_{j}(t_{0})}\right)\nonumber\\&-{1\over2}\sum_{i= j=1}^{3} {\dot{{\cal Z}}_{i}(t)\over {\cal Z}_{i}(t)}{\dot{{\cal Z}}_{j}(t_{0})\over {\cal Z}_{j}(t_{0})}+\frac{1}{2}\sum_{i=1}^{3} \left(  {\dot{{\cal Z}_{i}}(t)\over {\cal Z}_{i}(t)}\right)\biggr(\frac{\dot{\sigma}(t)}{\sigma(t)}+\frac{\dot{\sigma}(t_{0})}{\sigma(t_{0})}\biggr)\nonumber\\&+\frac{1}{2}\sum_{i=1}^{3} \left(  {\dot{{\cal Z}_{i}}(t_{0})\over {\cal Z}_{i}(t_{0})}\right)\biggr(\frac{\dot{\sigma}(t)}{\sigma(t)}+\frac{\dot{\sigma}(t_{0})}{\sigma(t_{0})}\biggr)+\frac{1}{2}\biggr(\frac{\dot{\sigma}(t)}{\sigma(t)}-\frac{\dot{\sigma}(t_{0})}{\sigma(t_{0})}\biggr)^2 +(\sigma(t)-\sigma(t_0))^2\biggr)\biggr]\nonumber\\
&\approx {(t-t_0)^2\over 4m^2t_0^4}.
\end{align}
 Analogously, we can also calculate the particle production for the massive
inflaton in single-mode $ STVS $. For this, using   the expectation values of $ \langle \Phi^2\rangle_{STVS} $ and $ \langle \Pi^2\rangle_{STVS} $ from Eqs. (23), (24) and $ \langle \Pi\Phi\rangle_{STVS} $, $ \langle\Phi \Pi\rangle_{STVS} $ can be calculated as

\begin{align}\label{eq5.6}
\langle\hat{\Pi}\hat{\Phi}\rangle_{STVS}
&={\cal Z}_{1}(t){\cal Z}_{2}(t){\cal Z}_{3}(t)\biggr[(\cosh^2\rho\ \cosh^2r_{s}-2A(\cos\theta)+\sinh^2\rho\ \sinh^2r_{s})\nonumber\\
&\times\dot{\Phi}(t_0){\Phi}^*(t_0)-(K)\dot{\Phi}^*(t_0)\Phi^*(t_0)-(L)\dot{\Phi}(t_0) \Phi(t_0)\nonumber\\&+(\sinh^2\rho\ \cosh^2r_{s}-2A(\cos\theta)+\cosh^2\rho\ \sinh^2r_{s})\nonumber\\
&\times\dot{\Phi}^*(t_0)\Phi(t_0)\biggr],
\end{align}
and
\begin{align}\label{eq5.7}
\langle\hat{\Phi}\hat{\Pi}\rangle_{STVS}
&={\cal Z}_{1}(t){\cal Z}_{2}(t){\cal Z}_{3}(t)\biggr[(\cosh^2\rho\ \cosh^2r_{s}-2A(\cos\theta)+\sinh^2\rho\ \sinh^2r_{s})\nonumber\\&\times \Phi(t_0)\dot{\Phi}^*(t_0)
-(K){\Phi}^*(t_0)\dot\Phi^*(t_0)-(L){\Phi}(t_0)\dot \Phi(t_0)\nonumber\\&+(\sinh^2\rho\ \cosh^2r_{s}-2A(\cos\theta)\nonumber\\&+\cosh^2\rho\ \sinh^2r_{s}){\Phi}^*(t_0)\dot\phi(t_0)\biggr],
\end{align}
Substituting Eqs.(23), (24), (48) and  (49) in (44), we can be calculated the number of particle produced in  single-mode $ STVS $ as

\begin{align}\label{eq5.8}
\langle{\hat{\cal N}}\rangle_{STVS}&=\biggr(\sinh^2\rho\ \cosh^2r_{s}-A(2\cos\theta)+\cosh^2\rho\ \sinh^2r_{s} +1\biggr)\nonumber\\&\times{\cal N}_0(t,t_0)+\sinh^2\rho\ \cosh^2r_{s}-A(2\cos\theta)
+\cosh^2\rho\ \sinh^2r_{s}\nonumber\\&-({\cal Z}_{1}(t){\cal Z}_{2}(t){\cal Z}_{3}(t))^2(K)E-({\cal Z}_{1}(t){\cal Z}_{2}(t){\cal Z}_{3}(t))^2(L)F,
\end{align}

where $ K $, $ L $ are given by Eqs.(25), (26) and

\begin{align}\label{eq5.9}
E & ={\exp(2i\int mdt_0)\over 4m^2({\cal Z}_{10}(t){\cal Z}_{20}(t){\cal Z}_{30}(t))^{2}\ t^2\ t^2_0}\left[{1\over
t^2_0}+{1\over t^2}-{2\over t t_{0}}-{2im\over t_0}+{2im\over t}\right], 
\end{align}
and
\begin{align}\label{eq5.10}
F 
&={\exp(-2i\int mdt_0)\over 4m^2({\cal Z}_{10}(t){\cal Z}_{20}(t){\cal Z}_{30}(t))^{2}\ t^2\ t^2_0}\left[{1\over
t^2_0} - {2\over tt_0}+{1\over t^2}+{2im\over t_0}-{2im\over t}\right].
\end{align}
In the same way, using the perturbative solution the number of created particles  in the limit  $mt_{0}$ , $mt>1$ from single-mode $  STVS$  can be calculated as [80] 
 
 \begin{align}\label{eq5.11}
{\cal N}_{STVS}(t, t_{0})\ \approx \ &\biggr(\sinh^2\rho \ \cosh^2r_{s}-A(2\cos\theta)+\cosh^2\rho\ \sinh^2r_{s}  +1\biggr)
 \frac{(t-t_{0})^2}{4m^2t_{0}^4}\nonumber\\
&+(\sinh^2\rho\ \cosh^2r_{s} - A(2\cos\theta)+\cosh^2\rho\ \sinh^2r_{s})\nonumber\\
&-2\biggr( B\cosh^2r_{s}\cos(2mt)-C( \sinh^2\rho +\cosh^2\rho )\nonumber\\&+B\sinh^2r_{s}\cos(2mt)\biggr) \frac{(t-t_{0})^2}{4m^2t_{0}^4}.
\end{align}

\section{Conclusions}

In this  research article, we have considered minimally coupling between  a massive inflaton and  the gravity and analyzed the behavior of massive inflaton field  in the anisotropically expending universe which is determined by the  Bianchi type-I  universe in the framework of $ SCTG $. The quantum and finite thermal effects of the massive inflaton was studied by representing the inflaton in  nonclassical thermal state. We particularly addressed the oscillatory regime of the massive  inflaton  in the single-mode $ STVS $ formalism. In this state formalisms, we obtained the approximate leading solution to the $ SCEE$ for the cosmic scale factor.
From Eq. (42)  it follows that ${\cal Z}_{11}(t)_{STVS}\sim t^{2\over3} $, ${\cal Z}_{21}(t)_{STVS}\sim t^{2\over3} $, $ {\cal Z}_{31}(t)_{STVS} \sim t^{2\over3} $.
Therefore, in the oscillating region of a massive inflaton in single-mode  $STVS$  formalism an approximate 
dominating solution to the  $ SCEE $ has 
equivalent power-law expansion for the matter-dominated universe $  t^{\frac{2}{3}} $ as  classical Einstein equation. Further,
the solution for one of the cosmic scale factors dependent on the initial value of the other cosmic
scale factors of other directions. Hence it can be concluded that, evolution of the cosmic
scale factors are related mutually. If we put
$ {\cal Z}_{1}(t) $ = $ {\cal Z}_{2}(t) $ = $ {\cal Z}_{3}(t) $=$ {\cal Z}(t) $, then the corresponding result reduces into the FRW model [62]. A damping process play an important role in the transformation from anisotropic to isotropic model. One of the
adequate damping process could be noticed due to the particle production in anisotropic
models [19]. Hence, particle production mechanism can bring isotropy in the Bianchi type-I anisotropic models.\vspace{1mm}

By the use of  $STVS$,  we also examined particle production of the oscillating  massive inflaton  in the framework of the $SCTG  $ in the context of the Bianchi
type-I cosmological model.   Using the single-mode $STVS$, we  computed the production of particles in the limit $m t_{0}, mt >1 $, here, we observed that produced particles depends on the associated thermal parameter ($ \rho$) and squeezing parameter ($ r_{s} $).  If we establish  $ \rho=0 $ in  Eq.(53) then we obtained the number of produced particles  in squeezed vacuum state which is totally match with  the results of  J. K. Kim and S. P. kim [81].  It is also noted that in the Bianchi type-I cosmological anisotropic model the coherently oscillating massive inflaton in single-mode $STVS$  suffering from particles production and produced particle show oscillations. After the inflation these oscillation of the produced particles is necessary to reheat the universe to hot again.   
 Therefore, the above-mentioned single-mode $ STVS $ can be  the possible quantum state with finite thermal effects in
which our universe could have been present during the matter-dominated era.\\

All authors have seen and approved the final version of the manuscript being submitted.
Authors have no conflict of interest to declare.
No data is use in this manuscript. No funding agencies included to fund this work.\\
\\
\textbf{References}\\
\bibliography{sn-bibliography}%

\end{document}